# Predictive genomics: A cancer hallmark network framework for predicting tumor clinical phenotypes using genome sequencing data

Edwin Wang[1,2,*], Naif Zaman[1,3], Shauna Mcgee[1,4], Jean-Sébastien Milanese[1,5], Ali Masoudi-Nejad[6] and Maureen O'Connor[1]

1. National Research Council Canada, Montreal, QC H4P 2R2, Canada

2. Center for Bioinformatics, McGill University, Montreal, QC H3G 0B1, Canada

3. Department of Anatomy and Cell Biology, McGill University, Montreal, QC H3A 2B2, Canada

4. Department of Experimental Medicine, McGill University, Montreal, QC H3A 1A3, Canada

5. Department of Medicine, Laval University, Quebec, QC G1V 0A6, Canada

6. Laboratory of Systems Biology and Bioinformatics (LBB), Institute of Biochemistry and Biophysics, University of Tehran, Tehran, Iran

*Corresponding author: Edwin Wang (edwin.wang@cnrc-nrc.gc.ca)

**Abstract**

Tumor genome sequencing leads to documenting thousands of DNA mutations and other genomic alterations. At present, these data cannot be analyzed adequately to aid in the understanding tumorigenesis and its evolution. Moreover, we have little insight into how to use these data to predict clinical phenotypes and tumor progression to better design patient treatment. To meet these challenges, we discuss a cancer hallmark network framework for modeling genome sequencing data to predict cancer clonal evolution and associated clinical phenotypes. The framework includes: (1) cancer hallmarks that can be represented by a few molecular/signaling networks. 'Network operational signatures' which represent gene regulatory logics/strengths enable to quantify state transitions and measures of hallmark traits. Thus, sets of genomic alterations which are associated with network operational signatures could be linked to the state/measure of hallmark traits. The network operational signature transforms genotypic data (i.e., genomic alterations) to regulatory phenotypic profiles (i.e., regulatory logics/strengths), to cellular phenotypic profiles (i.e., hallmark traits) which lead to clinical phenotypic profiles (i.e., a collection of hallmark traits). Furthermore, the framework considers regulatory logics of the hallmark networks under tumor evolutionary dynamics and therefore also includes: (2) a self-promoting positive feedback loop that is dominated by a genomic instability network and a cell survival/proliferation network is the main drivers of tumor clonal evolution. Surrounding tumor stroma and its host immune systems shape the evolutionary paths; (3) cell motility initiating metastasis is a byproduct of the above self-promoting loop activity during tumorigenesis; (4) an emerging hallmark network which triggers genome duplication dominates a feed-forward loop which in turn could act as a rate-limiting step for tumor formation; (5) mutations and other genomic alterations with specific patterns and tissue-specificity, which are driven by aging and other cancer-inducing agents.

This framework represents the logics of complex cancer biology as a myriad of phenotypic complexities governed by a limited set of underlying organizing principles. It therefore adds to our understanding of tumor evolution and tumorigenesis, and moreover, potential usefulness of predicting tumors' evolutionary paths and clinical phenotypes. Strategies of using this framework in conjunction with genome sequencing data in an attempt to predict personalized drug targets, drug resistance, and metastasis for a cancer patient, as well as cancer risks for a healthy individual are discussed. Accurate prediction of cancer clonal evolution and clinical phenotypes will have substantial impact on timely diagnosis, personalized management and prevention of cancer.

# 1 Introduction

Tumor genome sequencing has generated information on thousands of mutations and other genomic alterations. To date, more than 10,000 tumor genomes have been sequenced and as sequencing costs drop, many more genomes will be determined in the near future. Recently, Illumina released a high-throughput genetic sequencing machine, the HiSeq X Ten that can sequence a whole human genome for $1,000. New technologies such as Quantum Sequencing platform and Oxford Nanopore systems hold promise for further reducing genome sequencing cost in the future. This trend suggests that genome sequencing could be used as a diagnostic tool in clinical practice.

Tumor genome sequencing has cataloged many 'driver-mutating genes' and will continue to catalog more. However, the biological complexity of cancer combined with a vast amount of genome sequencing data presents a significant challenge on how to extract useful information and translate them into mechanistic understandings and predictions for cancer phenotypes [1], thus enabling management of cancer patient treatment in a more efficient manner. If one looks at a cancer cell genome, it is abundant with gene mutations, deletions and amplifications, chromosome gains and losses. Though seemingly random, there could in fact be patterns for mutations and chromosomal changes and by uncovering these patterns, we could in turn gain more insight into the mechanisms which drive cancer progression. By accessing the complete, precise genomic and clinical information from cancer patients, future advances will depend on exploiting the natural genetic complexity through uncovering key components of the cancer system. The genetic complexity observed can be described in a mathematical manner and models computed as a result. Moreover, the key components of cancer could be experimentally perturbed to test their characteristics. Ultimately, we could predict evolutionary path of the tumor clones and their phenotypes associated with progression, metastases and drug resistance.

In this review, we discuss a cancer hallmark network framework that can be used to model key components of cancer systems and then link mutant genotypes (i.e., mutations and other genomic alterations) to cellular and clinical genotypes. Using this framework, we illustrate strategies for prediction of cancer drug targets, probability of tumor recurrence, and cancer risks based on individual patient's genome sequencing profile. Predictions derived from cancer hallmark network-based modeling could ultimately be used in diagnosis and optimized patient management and prevention of cancer.

# 2 Cancer hallmarks and their networks

Although the biology of cancer is extremely complex, key traits have been revealed during the past decade. The complexity of cancer can be reduced and represented by a few distinctive and complementary capabilities ('cancer hallmarks') that enable tumor growth and metastasis dissemination. These hallmarks constitute an organizing principle that provides a logical framework for understanding the remarkable diversity of neoplastic diseases. In 2000, Weinberg and Hanahan proposed six cancer hallmarks [2]: (1) cancer cells self-stimulate their own growth; (2) they resist inhibitory signals that might prevent their growth; (3) they resist their own programmed cell death; (4) they stimulate the growth of blood vessels to supply nutrients to tumors; (5) they can multiply forever; and (6) they invade local tissue and spread to distant organs. These are the core common traits that govern the transformation of normal cells to cancer (malignant or tumor) cells. In 2011, Weinberg and Hanahan updated the cancer hallmarks by adding 4 more [3]: (7) abnormal metabolic pathways; (8) evading the immune system (escaping from immunosurveillance); (9) chromosome abnormalities and unstable DNA (genome instability); and (10) inflammation.

**2.1 Genome duplication is an emerging cancer hallmark**

Cancer hallmarks are evolving as we understand more about cancer. For example, from genomic point of view, the most striking characteristic of cancer genomes is extensive aneuploidy. Cancer genomes carry extremely high frequency of somatic copy number alterations, most of which are large-scale at chromosomal level. For example, several chromosomal arms can be amplified or deleted. It has been proposed that genome duplication could play a critical role for generating cancer aneuploidies and most likely be a rate-limiting step for tumor development [4, 5]. This assumption is based on the observations that genome duplication occurs only one time and routinely at the last round of gene amplification/deletion events during the transformation from normal to cancer cells [4, 5]. Aneuploidy is extensive in cancer [6]. We and others have shown that that a large fraction (~50%) of solid tumors have undergone genome duplication (Milanese et al. unpublished data; [7]. Therefore, we regard genome duplication as an emerging cancer hallmark trait which not only drives cancer aneuploidies but also a rate-limiting driving force during tumor formation. Genome duplication enables subtle changes in the activity of many different genes simultaneously and could facilitate the activation of several hallmark networks in one shot [4, 5]. In particular, the capabilities of interactions/regulations between hallmark networks could easily be acquired through genome duplication. Therefore, activation of genome duplication network could represent a "perfect storm" of extreme changes of genes in the cancer genome so that a cell could acquire a set of cancer hallmark traits at one time and then transform a 'slow-growing' cancer clone into 'fast-growing' one and therefore speed up tumor formation.

**2.2 Mapping of cancer hallmark traits onto cancer hallmark networks**

Over the past few years it has been argued that network or a systems approach should be adopted for modeling of cancer genomic data so as to further understand cancer biology and translate the information into clinical practice [1, 8, 9]. In the article of '*The Roadmap of Cancer Systems Biology*' [10], we proposed that a signaling network could be constructed for an individual tumor and modeling efforts could then be made ascribing cancer hallmark networks to that particular tumor. Substantial efforts have been made in this direction recently and have generated interesting results. For example, patient-specific whole signaling pathways have been constructed using a graphical model (PARADIGM) based on genomic alteration data and used for predicting drug targets [11]. We have developed an algorithm by modeling cancer hallmark gene modules and successfully identified highly robust cancer biomarkers [12]. Moreover, by the modeling of genomic alterations on the core cancer hallmark network – the cell survival and proliferation network - we are able to effectively predict (with 80% accuracy) breast cancer subtype-specific drug targets [13]. Theses examples are encouraging for modeling of tumor genomic data and complex cancer systems using a framework that consists of a set of hallmark networks representing underlying principles and mechanisms of tumorigenesis.

The mechanisms of cancer etiology attributed to signaling pathways of some cancer hallmarks are closely intertwined. Therefore, the hallmarks whose underlying signaling pathways are highly intertwined can be collected into one hallmark network. For example, the signaling processes of Hallmarks 1, 2 and 3 are highly interactive, which one can define as a cancer cell survival and proliferation network (for simplicity, the survival network). This survival network collects the interactions and signaling processes of the above three hallmarks. Using this principle, we mapped cancer hallmark networks onto hallmark traits. Finally, we represent the complexity of cancer using only ten hallmark networks (**Box 1**).

A cancer cell needs to acquire functional capabilities that allow it to survive, proliferate, disseminate and colonize (i.e., tumor metastasis) in distant organs. These functions are acquired in different tumor types via activating distinct hallmark networks and at various times during the course of multistep tumorigenesis. Among the networks, three of them are core networks and critical for these multistep processes. The mutation network, which represents Hallmark 9 (i.e., genome instability), not only is the master driver for tumorigenesis but also orchestrate hallmark capabilities. Acquisition of other hallmark traits mainly depends on the activation of the mutation network. The survival network represents the fundamental activities (i.e., survival and proliferation) of a cancer cell, while EMT network representing Hallmark 6 carries out the initial key step of tumor metastases. Details for the hallmark network relationships in a context of tumor evolutionary dynamics have been illustrated in **Fig 1a**.

**2.3 Quantifying cancer hallmark traits and networks to link mutant genotype to regulatory, molecular and cellular phenotypes of cancer cells**

Although cancer hallmark concepts have been introduced over 10 years, they are largely described in text. Aside from traits such as cell proliferation and EMT, most of the hallmark traits have no quantification measures proposed. To model cancer hallmark traits and networks, it is necessary to quantify not only hallmark networks but also hallmark traits. We could then link network measures to hallmark trait measures using mathematical language.

Network measures can be quantified at two levels: operational (i.e., regulatory) and molecular (i.e., functional) levels. The topological structure of a molecular network can be seen as computer hardware, while gene/protein regulations/interactions on the network can be seen as software programs which run on computers. Different cancer cells/clones could run distinct software programs. For example, different cancer cells have different sets of genomic alterations, which could trigger distinct operational (regulatory) programs on the network. In this context, nodes and links in the network could be weighted, and certain scoring functions (representing strengths of gene/protein regulations) could be developed to represent gene regulatory logics/strengths on networks. We define a 'network operational signature' which is represented by a regulatory profile in which a set of genes form a set of regulatory circuits encoding gene regulatory relationships and strengths. Gene expression has been regarded as a molecular phenotype. Similarly, gene/protein regulatory (including signaling, protein modifications, and other regulatory relations such as non-coding RNA regulations) logics/strengths can be seen as a 'regulatory phenotype'. A same set of genomic alterations could have distinct regulatory logics/strengths (i.e., regulatory phenotypes) on distinct molecular networks, and therefore, could lead to distinct cellular phenotypes in different cells. A gene expression signature representing specific molecular phenotypes can be linked to cellular or clinical phenotypes, therefore, we propose that network operational signatures representing specific regulatory phenotypes can be linked to cellular or clinical phenotypes as well. Details of network operational signatures associated with cancer hallmarks are described in Section 3.1. At the molecular level, we define network measures using gene expression or protein phosphorylation signatures, in which both gene expression and phosphorylation are treated as molecular phenotypes of cells. We propose phosphorylation signatures as a measuring feature due to our belief that cell signaling networks are critical for tumorigenesis [4, 5]. A molecular signature contains a list of genes and their values of either gene expression or phosphorylation, while a network operational signature contains a set of genes encoding their regulatory relationships and strengths.

Cancer hallmark traits can be quantified based on the cell biology of the hallmarks. For example, cell survival could be measured by the cell proliferation rate or by the proliferation marker Ki-67; genome instability could be measured by the mutations density or deletion/amplification density per Mb chromosome; cancer dissemination could be measured by the number of cancer cells circuited in blood (volume) and the primary tumor (volume) and so on. With the quantification of the networks and their corresponding hallmark traits, we could model their relationships using computational approaches. Previously, we have modeled cancer hallmarks using gene expression signatures (i.e., functional gene modules of cancer hallmark networks) and these molecular signatures effectively predicted tumor metastasis events (Li 2010). Recently, we also developed an algorithm to identify network operational signatures of cancer cell survival networks that link to cancer cell drug targets (Tibichi et al., unpublished data). Using this algorithm, we correctly predicted which genes are essential to cell survival by providing the genomic alterations of an individual cancer cell which in turn serve as drug targets.

**3 Principles of the cancer hallmark network framework**

Hallmark networks possess distinctive and complementary capabilities that enable tumor growth and metastasis, which constitute organizing principles that provide a logical framework for understanding the remarkable diversity of the diseases. We hypothesize that a collection these hallmark networks orchestrates and drives tumorigenesis, and thus forms a framework for network-based modeling and predictions of cancer phenotypes and clonal evolutionary paths (e.g., we could predict tumor clonal evolving several-steps-ahead). In this framework, higher-order rules between hallmark networks in the context of tumor evolutionary dynamics exist (Fig 1). These rules help us to prioritize the networks that are more suitable for the modeling of distinct stages of tumorigenesis. By developing this framework, we foresee cancer research as an increasingly logical science, in which a myriad of phenotypic complexities are manifestations of a small set of underlying organizing principles.

**3.1 Transformation of genomic alternations into network regulatory phenotypes which quantify cellular phenotypes such as hallmark traits**

From a systems biology perspective, each hallmark network has a critical transition threshold (i.e., tipping point) at which the system shifts abruptly from one state to another [4, 5]. This tipping point can be quantified by a network operational signature and furthermore, it can be reflected in hallmark traits. For example, a critical transition threshold for a survival network is marked by the significantly different states between normal and cancer cells. In this context, a network operational signature could specifically quantify the sharp transition between States 0 and 1 (for example, for the cell survival hallmark, 0 represents the normal cell state while 1 represents the cancer cell state). In Stage 1, some hallmark traits have their own measurements, such as proliferation rate, which is different between cancer cells. A network operational signature can specifically quantify the measures of hallmark traits. In this context, we propose that network operational signatures of a hallmark network could be identified to quantify both a transition between States 0 and 1 and the measure of State 1 of hallmark traits.

A series of network analyses of cancer genes showed that genes for each hallmark trait are most likely to be enriched in subnetworks or network communities [14-20], suggesting that sets of genes work together to accomplish the task. It has been proposed that there exist recurrent positive feedback network motifs, or functional modules, during the sharp transitions of States 0 and 1 [4, 5]. A network operational signature could be enriched with regulatory relations and strengths of the genes within one or few network motifs/network communities. The core hypothesis is that multiple cancer-driving factors, such as genomic alterations, could have thousands of combinations (ways) to trigger a 'state transition' of a hallmark network. However, to trigger a state transition they must go beyond the critical transition threshold which can be quantified by a network operational signature. Similarly, there are many ways (combinations of cancer-driving factors) in a hallmark network could drive different measures of a hallmark trait, but only a limited number of network operational signatures can quantify them. Importantly, by developing appropriate scoring functions, we could transform the combinations of cancer-driving factors (i.e., genomic alterations) into network operational signatures which could directly quantify hallmark traits. This modeling approach quantitatively links genotypes (genomic alterations) to regulatory phenotypes (regulatory logics/strengths), to cellular phenotypes (hallmark traits) and finally to clinical phenotypes (a collection of hallmark traits).

State transitions in network modeling are often described by activation/inactivation status of 'output nodes/genes' or 'sink' of networks [15]. However, in many cases, it is unclear which genes are the output nodes for a biological process and furthermore, output nodes of the networks could be different between cancer cells. It is a very challenging task to define the output nodes for the networks of cancer cells. Therefore, network operational signatures provide an effective means which overcomes the problems of defining output nodes of a functional network.

### 3.2 Cancer cell evolution is mainly driven by a self-promoting positive feedback loop of hallmark networks

The evolutionary process of cancer cells is highly dynamic, however, the outcome of evolution and selection will finally lead to favor hyper-proliferation and fast-growing cancer cells. The biology of a tumor cannot be understood by simply enumerating the cancer hallmark traits or networks, but instead must consider the interactions/regulations of these hallmark networks in terms of evolutionary dynamics. Therefore, we propose that the major evolutionary driving force of a cancer cell comes from a self-promoting positive feedback loop in which mutation network and survival network are dominant components (**Fig 1b**). In this loop, mutation network is a master activator/regulator, which is a driving engine for genomic alterations, while survival network is the final regulated target. This survival network, which continually sends positive feedback (i.e., selecting cancer cells with more fitness) to the mutation network, drives the evolutionary path of cancer cells. Mutation network induces high instability of the genome leading to an increase in genomic alteration occurring in the cell. New genomic alterations could be occurred in some components (i.e., genes/proteins) of the mutation network, which in turn could give the network higher efficient, thereby inducing a higher level of genome instability. The mutation network therefore has a self-regulating function.

Genomic alterations induced by a mutation network could occur in the components of other hallmark networks so that other hallmark traits could be acquired. Cancer cell populations are much like an ecosystem in which fast-growing cancer cells have higher competitive power than slow-growing ones. Fast-growing cancer cells will be selected to become a dominate subpopulation within a tumor and therefore a highly efficient survival network will be selected. We must be made aware that the networks in the self-promoting loop are in the same cell. If a highly efficient survival network is selected, the mutation network and other networks in the loop are also selected. This selection process is equivalent to sending a positive feedback to these networks by the survival network. Combined with the knowledge that the mutation network can be self-regulated, we propose that the loop is a self-promoting positive feedback loop, dominated by the networks of mutation and survival and refer to it as self-promoting loop (**Fig 1b**).

The two main classes of cancer genes are oncogenes and tumor suppressor genes. In general, oncogenes are often mutated or amplified in positive gene circuits, whereas tumor suppressor genes are often mutated, deleted or methylated in negative gene circuits [16]. Negative-feedback gene circuits that normally operate to dampen various types of signaling ensure homeostatic regulation of the flux of signals coursing through the intracellular circuitry [15, 21]. Defects in these circuits are capable of enhancing proliferate signaling [15]. Based on the model of the self-promoting positive feedback loop of hallmark networks, we expect that the 'destination' of tumor cells is to become hyperproliferative in nature by self-promoting their genomes to have a higher hypermutation rate (i.e, a highly efficient mutation network). Therefore, the outcome of this self-promoting process will lead more tumor suppressor genes becoming mutated/deleted and more oncogenes becoming mutated /amplified. This reasoning is supported by a recent survey that the distribution of oncogenes and tumor suppressor genes is correlated with copy number alterations (i.e., amplifications and deletions) of chromosome arms [6].

It is clear that the self-promoting loop mainly drives cancer cell evolution. However, intercellular systems could shape the path of cancer evolution. Stroma-network (e.g., inflammation and tumor stromal immune systems) could send signals to the networks of the self-promoting loop, and furthermore, tumor host immune system forms selection forces for cancer cells. These hallmark systems also play an important role in shaping cancer evolutionary path to drive tumorigenesis (**Fig 1a**).

### 3.3 A genome duplication network activated feedforward loop is a rate-limiting step for tumor formation

Genome duplication appears to be a rate-limiting step for tumor formation in many epithelial-origin tumor samples [4, 5]. A sharp transition between a slow-growing clone to a fast-growing clone could be governed by a genome duplication event yet thus far, it is unclear how genome duplication is triggered. Interactions between the mutation network of a cancer cell and its host systems such as tumor stroma-network may activate genome duplication network thereby trigger a genome duplication event in that cancer cell. This event often drives a large number of amplifications and deletions to occur simultaneously and in turn other hallmark networks, such as survival network, could quickly reach high efficiency. Therefore, the selection process after a genome duplication event is governed by a feedforward loop (**Fig 1c**) in which a genome duplication network is the master regulator, the dedifferentiation network, angiogenesis-inducing network and immune-escaping network are the secondary regulators, and survival network as the regulated target. If this feedforward loop is more efficient than the step-wise self-promoting positive feedback loop, the number of amplified and deleted genes in tumors which experienced genome duplication could be significantly smaller than that of the tumors which had not experienced genome duplication. Indeed, we found that this is true in our analysis of a few hundreds of tumor genomes (Milanese et al. unpublished materials).

### 3.4 Cancer cell motility initiating metastasis is a byproduct of the cancer hallmark self-promoting loop activity

Tumor recurrence and metastasis are the leading cause of cancer mortality. Therapies for recurrent disease may fail, at least in part, because the genomic alterations driving the growth of recurrences are distinct from those in the initial tumor. Therefore, understanding how metastasis occurs is critical. The first step of metastasis is a cancer cell that acquires motility governed by an epithelial-mesenchymal transition (EMT) network.

As the activity of mutation network gets higher, more genomic altercations will occur in the genome. If certain genomic altercations hit EMT network components, a transition between States 0 and 1 could be triggered (i.e., 0 represents a cancer cell without motility capability, while 1 represents a cancer cell acquired motility capability). Once a cancer cell acquires the EMT trait, it starts to move out of the tumor, then circulates in the host systems and finally colonizes in distant organs. In this context, cancer cell motility is not a trait which has been selected for, but it is mainly driven by the mutation network and the self-promoting loop. If this is true, we expect that EMT-driven cell motility could be acquired by cancer cells at any time during its evolution, including the slow-growing stage (i.e., before primary tumor formation, or from primary tumors). However, generally if EMT occurs in fast-growing cells, these cells could have higher chance to undergo metastasis. Traditionally it is believed that metastasis occurs late in tumor formation (i.e., dissemination of tumor cells in late stage of tumor progression). However, a number of recent studies indicate that early dissemination of tumor cells occur before tumor formation [3].

### 3.5 Mutations and other genomic alterations have specific patterns and tissue-specificity

For years it has been known that mutational signatures exist in cancer. For example, both UV light and tobacco-smoking produce very specific signatures in a person's genome. It is assumed that each mutational process leaves a particular pattern of mutations, an imprint or signature, in the genomes of cancers it causes. Using an algorithm previously developed for searching mutational patterns [22], investigators analyzed more than 7,000 tumor genomes representing 30 most common cancer types and found sets of mutational signatures in tumors [23]. They found that each tumor type contains mutational signatures derived from at least 2 different and distinct factors such as aging, smoking, UV and so on. This means that we can now study those patterns that drive cancer and furthermore, it suggests that cancer development, progression and metastases can be predicted.

If agent-derived mutational signatures are evenly distributed across all tissues, it is difficult to explain why only some agents induce certain types of tumors. With the assumption that a mutational signature could occur in distinct tissues and tumor types, we carefully checked the results obtained from a recent survey of mutational signatures among 30 tumor types [23] and found our hypothesis to be valid. An agent-derived mutational signature can indeed be tissue-dependent. For example, BRCA1/2-mutation-derived mutational signatures dominantly occur in the tumors of breast, ovary and pancreas [23]. Aging-derived mutational signatures can be both age- and tissue-specific-dependent. For example, aging-derived Siganture1A predominantly occurs in AML and other 6 tumor types [23]. In addition, an agent could induce two or more different mutational signature profiles which could occur in distinct tissues. For example,

APOBEC is able to induce at least two different mutational signatures: one occurs in 16 tumor types while the other occurs in 2 other tumor types [23]. Furthermore, a recent study showed that tissue-specific mtDNA mutation signatures are common, even in normal tissues, and also showed that most of these signatures are aging-driven [24]. This study underscores that human DNA changes or mutates in patterns rather than randomly.

These mutational signatures will be extremely useful for modeling genome sequencing data using the hallmark network-based framework. Therefore, it is critical to survey more genomes with different ages and from different tissues, to create a compendium of mutational signatures triggered by the aging process. In addition, a similar effort should be taken in collecting mutational signatures derived from cancer-inducing factors (including chemotherapy drugs and therapy-based radiations) in different tissues and cancer types.

**4 Strategies for constructing predictive models using a hallmark network framework**

Tumor progression can be portrayed as a succession of clonal expansions, each of which is triggered by the interactions of a set of hallmark networks [4, 5]. Starting from a normal cell to a cancer clone, new mutations during clonal expansion will build on the existing mutation profiles of previous clones. Therefore, a hallmark network can be updated and examined to check if the newly added mutations could work together with the 'pre-existing mutations' of the network to trigger a state transition or modulate the measures of a hallmark trait. Mutational patterns can be treated as key factors in the modeling experiments. For example, age-associated mutational patterns are major driving force for tumorigenesis. Different age groups have distinct mutational patterns, but they do not exist across all the tissue types. If we know the mutational patterns driven by aging or other factors, we could simulate the mutations of the proteins (or the regulatory or post-modification elements of the proteins such as transcription factor binding sites and phosphorylation motifs), which contain sequences matching that mutational pattern (**Fig 2a**). From there we could predict which proteins on a network could become mutated and whether this mutation works together with the 'pre-existing mutations' of the network to trigger the activation of a hallmark network by comparing with network operational signatures (**Fig 2a**). Network operational signatures could be identified from the gene regulatory profiles, which are generated from the network of normal and tumor samples. By doing so, cancer clonal evolution becomes largely predictable. The predictability of tumor clonal evolution means that we could foresee clonal evolutionary paths and then design cancer therapies to effectively conduct individualized cancer treatment.

More than 10,000 tumor genomes have been sequenced and this effort is still ongoing. It is expected that more tumor genomes and their matched gremlin genomes will be sequenced in the near future. Using these data, we can obtain network operational signatures for quantifying state transitions and measures for hallmark traits. Using these signatures and the hallmark network framework we can construct models that predict how the tumor will evolve and the risk potentials for healthy people who have certain germline mutations which could induce cancers.

**4.1 Predictive models for a cancer patient**

When a patient has a primary tumor only (i.e., metastasis has not been detected yet), it would be interesting to predict: 1) the possibility of tumor recurrence and metastases; 2) which drugs could be applied; and 3) the potential consequences after applying a specific drug. This information could aid in the design of therapies of appropriate first line of treatment and avoiding drug resistance derived from a specific drug treatment.

There are at least three key steps in the transformation of a normal cell to a metastatic cell (i.e., from a normal to a cancer cell; from a cancer cell to a circulating cell; and then to a metastatic cell which has been colonized in another organ). In this process, survival networks could be undergoing at least three key rounds of rewiring so that the drug targets for these three clinical phenotypes could be different. It would be useful to predict drug targets not only for the primary tumor, but also for the circulating cells so that circulating cancer cells could also get treated. As mentioned above, modeling of survival networks could predict drug targets for individual cancer cells. The survival signaling network can be constructed based on genome-wide RNAi knockdown data. Previously, we constructed breast cancer subtype-specific survival networks which were used for predicting subtype-specific drug targets with 80% accuracy (Zaman 2013). Recently, we developed network operational signatures of survival networks that enabled prediction of drug targets specifically for a single cancer cell/clone with 90% accuracy (Tibiche et al., manuscript in preparation). This tool allows predicting drug targets for each clone within a tumor based on their genomic alteration profiles. The genomic alternation profiles of primary tumor clones and the circulating cancer cells can be used for predicting their drug targets.

Currently, patients are often treated with generic cancer chemotherapeutic drugs which are often toxic and may also induce new mutations to tumor cells. To predict potential drug resistance, we could apply a specific drug to determine the mutational patterns driven by that drug on survival networks. For example, a drug can be applied to different cancer cell lines, followed by genome sequencing to identify mutational patterns derived from that specific drug treatment. If the mutational pattern driven by a specific drug that is to be applied to a patient is known, we could simulate the mutations of the proteins (whose sequences match the mutational pattern, of the current survival networks of the tumor clones) to predict the probability of drug resistance and help in making clinical decisions whether a drug should be applied to that patient. It should be noted that for that same drug, the probability of drug resistance varies between different patients (i.e., their survival networks have distinct pre-existing genomic alteration profiles).

To predict metastasis of a primary tumor, it is possible to simulate protein mutations of the EMT network based on mutational patterns learned from that tumor combined with aging-associated mutational patterns. The same method could apply to the mutation network and the immune-escaping network for the circulation and colonization processes of a cancer cell. In addition, host immune repertoire deep sequencing data could be integrated to examine the probability of tumor recurrence and metastasis.

The collection of these predictions could help in understanding of the mechanisms by which invading cells give rise to recurrent tumors and the effect of adjuvant therapeutics have on their evolution which will facilitate the development of new strategies to delay or prevent recurrence and malignant tumor progression.

**4.2 Predictive models for cancer risk for healthy individuals**

At the moment, cancer predisposition genes in germline cells are the major genetic factors to be used for evaluating cancer risks for a healthy person. To date, ~114 cancer predisposition genes have been identified [25]. The assumption is that if a cancer is detected early, treatment and survival will be improved. Preventative measures can be taken, such as the surgical removal of the at-risk tissue, however, it is preferable to conduct chemoprevention for persons at-risk. For example, taking an aspirin daily significantly reduced the risk of colorectal cancer for men who have mismatch repair gene mutation in germline cells [26]. However, it is unwise to use a single cancer predisposition gene to judge all cancer risks and we propose using the cancer hallmark framework (**Fig 2b**) could more effectively evaluate cancer risks based on a person's genomic profiles derived from blood cells.

We will consider the germline and aging mutational profiles, as well as potential environmental factors for inducing cancers in the framework. The self-promoting feedback loop (**Fig 1b**) containing both the mutation network and survival network could provide a window of opportunity for mitigating or preventing cancer. As shown in **Fig 2b**, hallmark networks will be first mapped using the mutations from the germline genomic profile, followed by other potential mutations based on the mutational patterns derived from aging and other potential cancer inducing agents (e.g., if a person is a heavy smoker, mutational patterns triggered by smoking will be projected). Network operational signatures can be used for predicting if and when the person could develop cancer. In this model, aging is considered as a major driving force for tumorigenesis: for example, nearly half of cancer patients are diagnosed after 65 years of age. Moreover, genome-wide age-related DNA methylation changes in blood and other tissues related to histone modification, expression and cancer, revealed the existence of age-specific DNA methylation genes [27], which suggests that age-specific mutation patterns exists. Furthermore, age-associated mutational patterns are tissue-specific. Given this strategy, the incidence and timing of malignant tumor progression can potentially be predictable. Because this prediction is based on hallmark networks in which gene regulatory profiles can be generated by mutations, it is possible to predict potential drug targets using the regulatory profiles. Therefore, it is possible to apply drugs for personalized cancer prevention.

**5 Challenges**
Our framework consists of a set of cancer hallmark networks. However, at the moment only a few hallmark networks are relatively rich in information, such as cell survival, mutation (i.e., genome instability) and EMT networks. Comprehensive information for other hallmark networks is still not available: a reasonably complete stroma-network is still not available as the great majority of signaling molecules and pathways remain to be identified. This reality poses challenges in using this framework to explore the interactions between stoma and cancer cells in orchestrating malignant progression.

Mutational patterns are critical factors in constructing predictive models, however, mutational patterns derived from aging, chemo and popular germline mutations have been not extensively catalogued. It is therefore important to gather a comprehensive compendium of mutational patterns for the understanding of cancer development and aid in constructing predictive models. Finally, it is essential to develop optimized network scoring functions for network modeling and network operational signatures for hallmark networks.

## References

[1] E. Wang, Understanding genomic alterations in cancer genomes using an integrative network approach. Cancer Lett. 340 (2013) 261-269.

[2] D. Hanahan and R.A. Weinberg, The Hallmarks of Cancer. Cell 100 (2000) 57-70.

[3] D. Hanahan and R.A. Weinberg, Hallmarks of cancer: the next generation. Cell 144 (2011) 646-674.

[4] E. Wang, J. Zou, N. Zaman, L.K. Beitel, M. Trifiro, and M. Paliouras, Cancer systems biology in the genome sequencing era: Part 1, dissecting and modeling of tumor clones and their networks. Semin. Cancer Biol. 23 (2013) 279-285.

[5] E. Wang, J. Zou, N. Zaman, L.K. Beitel, M. Trifiro, and M. Paliouras, Cancer systems biology in the genome sequencing era: Part 2, evolutionary dynamics of tumor clonal networks and drug resistance. Semin. Cancer Biol. 23 (2013) 286-292.

[6] T. Davoli, A.W. Xu, K.E. Mengwasser, L.M. Sack, J.C. Yoon, P.J. Park, and S.J. Elledge, Cumulative haploinsufficiency and triplosensitivity drive aneuploidy patterns and shape the cancer genome. Cell 155 (2013) 948-962.

[7] S.L. Carter, K. Cibulskis, E. Helman, A. McKenna, H. Shen, T. Zack, P.W. Laird, R.C. Onofrio, W. Winckler, B.A. Weir, R. Beroukhim, D. Pellman, D.A. Levine, E.S. Lander, M. Meyerson, and G. Getz, Absolute quantification of somatic DNA alterations in human cancer. Nat. Biotechnol. 30 (2012) 413-421.

[8] J. Shrager and J.M. Tenenbaum, Rapid learning for precision oncology. Nat. Rev. Clin. Oncol. 11 (2014) 109-118.

[9] M.B. Yaffe, The scientific drunk and the lamppost: massive sequencing efforts in cancer discovery and treatment. Sci. Signal. 6 (2013) e13.

[10] E. Wang, (2010) Cancer systems biology, CRC Press, Boca Raton, FL.

[11] C.J. Vaske, S.C. Benz, J.Z. Sanborn, D. Earl, C. Szeto, J. Zhu, D. Haussler, and J.M. Stuart, Inference of patient-specific pathway activities from multi-dimensional cancer genomics data using PARADIGM. Bioinformatics. 26 (2010) i237-i245.

[12] J. Li, A.E. Lenferink, Y. Deng, C. Collins, Q. Cui, E.O. Purisima, M.D. O'Connor-McCourt, and E. Wang, Identification of high-quality cancer prognostic markers and metastasis network modules. Nat. Commun. 1 (2010) 34.

[13] N. Zaman, L. Li, M.L. Jaramillo, Z. Sun, C. Tibiche, M. Banville, C. Collins, M. Trifiro, M. Paliouras, A. Nantel, M. O'Connor-McCourt, and E. Wang, Signaling network assessment of mutations and copy number variations predict breast cancer subtype-specific drug targets. Cell Rep. 5 (2013) 216-223.

[14] A. Awan, H. Bari, F. Yan, S. Moksong, S. Yang, S. Chowdhury, Q. Cui, Z. Yu, E.O. Purisima, and E. Wang, Regulatory network motifs and hotspots of cancer genes in a mammalian cellular signalling network. IET. Syst. Biol. 1 (2007) 292-297.

[15] M. Cloutier and E. Wang, Dynamic modeling and analysis of cancer cellular network motifs. Integr. Biol. (Camb. ) 3 (2011) 724-732.

[16] Q. Cui, Y. Ma, M. Jaramillo, H. Bari, A. Awan, S. Yang, S. Zhang, L. Liu, M. Lu, M. O'Connor-McCourt, E.O. Purisima, and E. Wang, A map of human cancer signaling. Mol. Syst. Biol. 3 (2007) 152.

[17] C. Fu, J. Li, and E. Wang, Signaling network analysis of ubiquitin-mediated proteins suggests correlations between the 26S proteasome and tumor progression. Mol. Biosyst. 5 (2009) 1809-1816.

[18] L. Li, C. Tibiche, C. Fu, T. Kaneko, M.F. Moran, M.R. Schiller, S.S. Li, and E. Wang, The human phosphotyrosine signaling network: Evolution and hotspots of hijacking in cancer. Genome Res. 22 (2012) 1222-1230.

[19] M. Paliouras, N. Zaman, R. Lumbroso, L. Kapogeorgakis, L.K. Beitel, E. Wang, and M. Trifiro, Dynamic rewiring of the androgen receptor protein interaction network correlates with prostate cancer clinical outcomes. Integr. Biol. (Camb. ) 3 (2011) 1020-1032.

[20] E. Wang, A. Lenferink, and M. O'Connor-McCourt, Cancer systems biology: exploring cancer-associated genes on cellular networks. Cell Mol. Life Sci. 64 (2007) 1752-1762.

[21] I.E. Wertz and V.M. Dixit, Regulation of death receptor signaling by the ubiquitin system. Cell Death. Differ. 17 (2010) 14-24.

[22] L.B. Alexandrov, S. Nik-Zainal, D.C. Wedge, P.J. Campbell, and M.R. Stratton, Deciphering signatures of mutational processes operative in human cancer. Cell Rep. 3 (2013) 246-259.

[23] L.B. Alexandrov, S. Nik-Zainal, D.C. Wedge, S.A. Aparicio, S. Behjati, A.V. Biankin, G.R. Bignell, N. Bolli, A. Borg, A.L. Borresen-Dale, S. Boyault, B. Burkhardt, A.P. Butler, C. Caldas, H.R. Davies, C. Desmedt, R. Eils, J.E. Eyfjord, J.A. Foekens, M. Greaves, F. Hosoda, B. Hutter, T. Ilicic, S. Imbeaud, M. Imielinski, N. Jager, D.T. Jones, D. Jones, S. Knappskog, M. Kool, S.R. Lakhani, C. Lopez-Otin, S. Martin, N.C. Munshi, H. Nakamura, P.A. Northcott, M. Pajic, E. Papaemmanuil, A. Paradiso, J.V. Pearson, X.S. Puente, K. Raine, M. Ramakrishna, A.L. Richardson, J. Richter, P. Rosenstiel, M. Schlesner, T.N. Schumacher, P.N. Span, J.W. Teague, Y. Totoki, A.N. Tutt, R. Valdes-Mas, M.M. van Buuren, '. van, V, A. Vincent-Salomon, N. Waddell, L.R. Yates, J. Zucman-Rossi, P.A. Futreal, U. McDermott, P. Lichter, M. Meyerson, S.M. Grimmond, R. Siebert, E. Campo, T. Shibata, S.M. Pfister, P.J. Campbell, and M.R. Stratton, Signatures of mutational processes in human cancer. Nature 500 (2013) 415-421.

[24] D.C. Samuels, C. Li, B. Li, Z. Song, E. Torstenson, C.H. Boyd, A. Rokas, T.A. Thornton-Wells, J.H. Moore, T.M. Hughes, R.D. Hoffman, J.L. Haines, D.G. Murdock, D.P. Mortlock, and S.M. Williams, Recurrent tissue-specific mtDNA mutations are common in humans. PLoS. Genet. 9 (2013) e1003929.

[25] N. Rahman, Realizing the promise of cancer predisposition genes. Nature 505 (2014) 302-308.

[26] J. Burn, J.C. Mathers, and D.T. Bishop, Chemoprevention in Lynch syndrome. Fam. Cancer 12 (2013) 707-718.

[27] Z. Xu and J.A. Taylor, Genome-wide age-related DNA methylation changes in blood and other tissues relate to histone modification, expression and cancer. Carcinogenesis 35 (2014) 356-364.

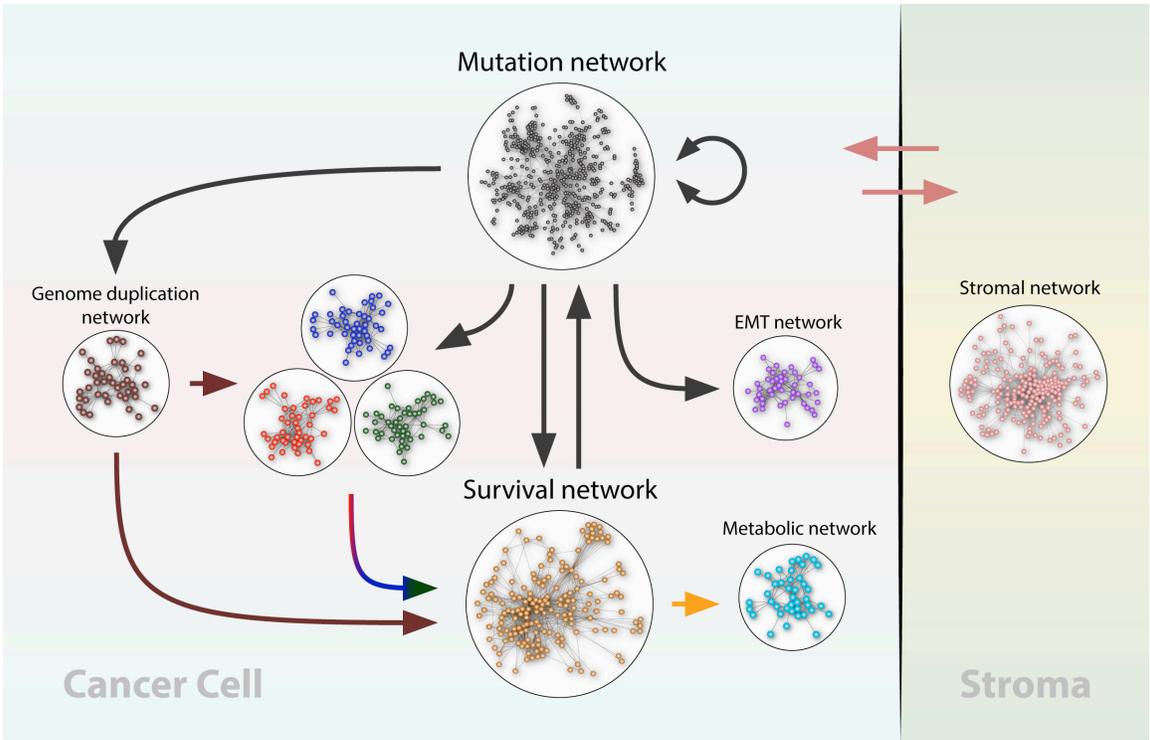

**1a**

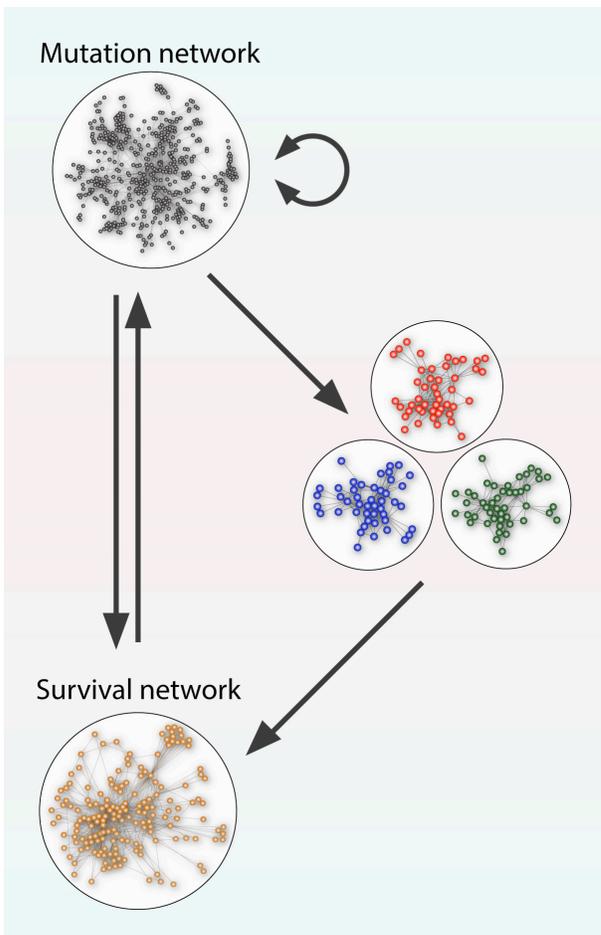

**1b**

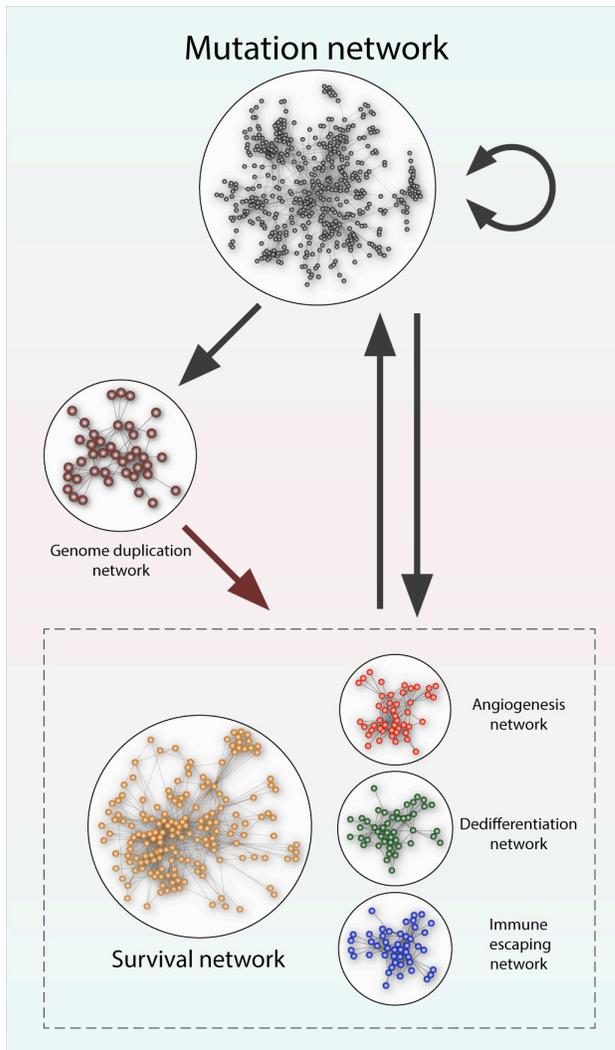

1c

**Fig 1. Cancer hallmark networks and their relationships in the context of tumor clonal evolutionary dynamics. Fig 1a** represents the regulatory relationships of cancer hallmark networks. **Fig 1b** represents a self-promoting positive feedback loop which is the dominantly driving force for cancer clonal evolution. In this loop, mutation network is a master activator/regulator, which is a driving engine for genomic alterations, while survival network is the final regulated target. This survival network, which continually sends positive feedback to the mutation network, drives the evolutionary path of cancer cells. **Fig 1c** represents a feedforward loop in which a genome duplication network is the master regulator, the dedifferentiation network, angiogenesis-inducing network and immune-escaping network are the secondary regulators, and survival network as the regulated target. The hallmarks represented by each network have been explained in Box 1. Networks in blue, red and green represent immune-escaping network, angiogenesis-inducing network, and dedifferentiation network, respectively. Line with arrows represents an inducing (positive) or interaction relationship, while open circle with double arrows represents self-regulation.

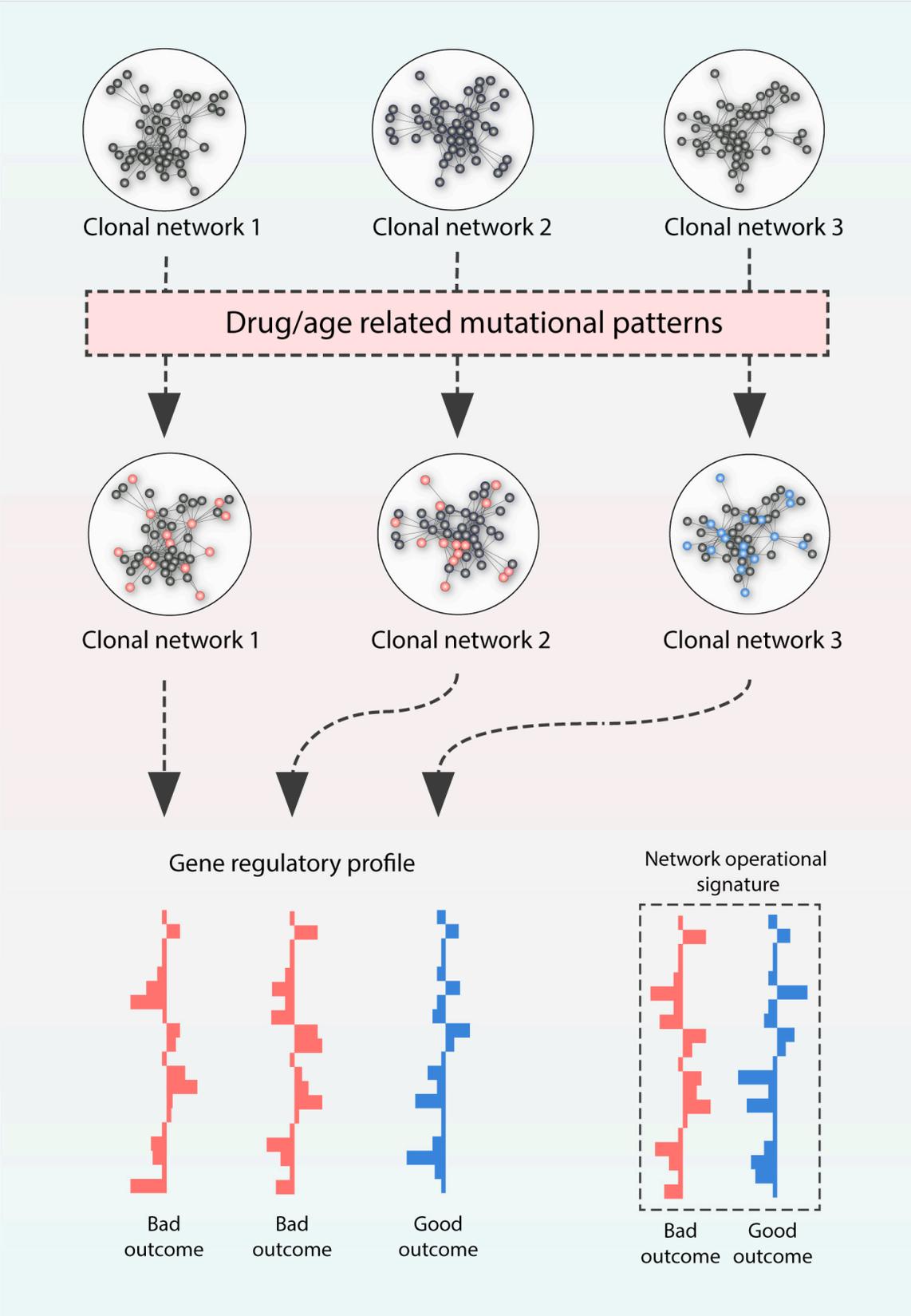

**2a**

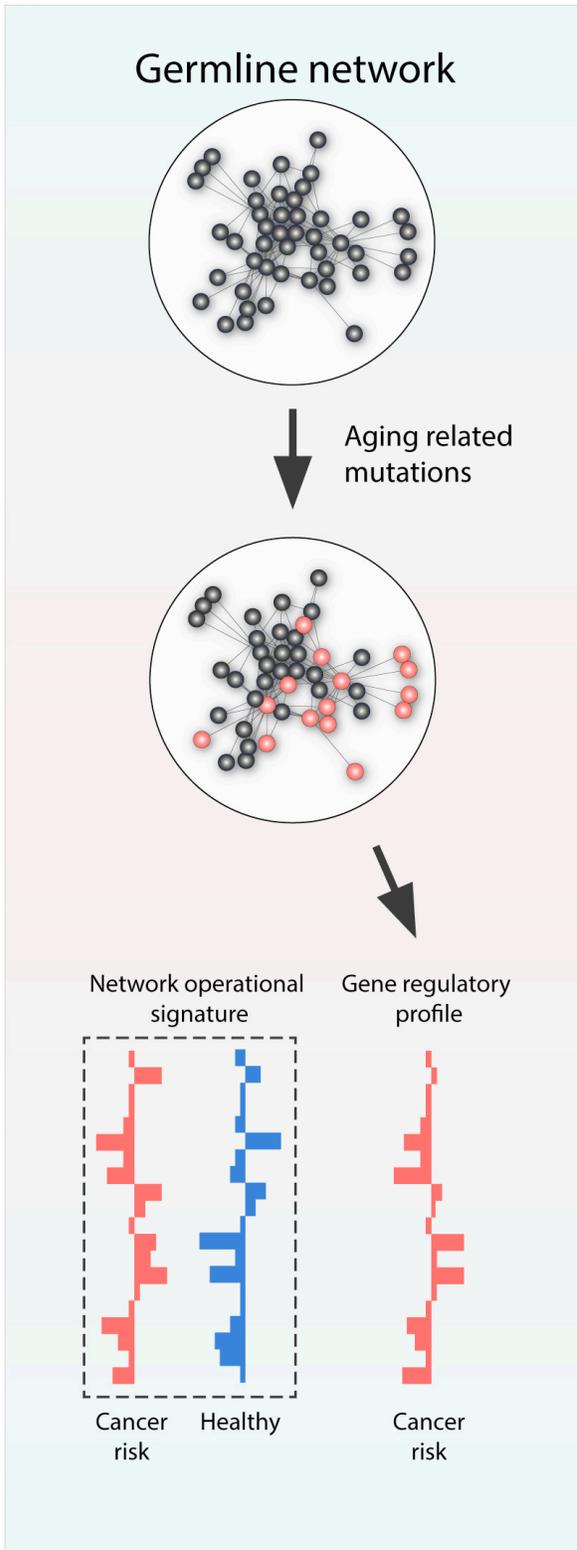

**2b**

**Fig 2. Constructing predictive network models using the cancer hallmark network framework.** Based on the relationships and functions of the hallmark networks, distinct networks could be modeled, for example, for predicting drug targets, survival network could be modeled; for predicting cancer risk, mutation network and survival network could be modeled; for predicting cancer recurrence, EMT network and stroma-network could be modeled. For each predictive model, network operational signatures, mutational patterns and so on will be used. New mutations based on mutational patterns of cancer-inducing agents could be projected onto the networks. Gene regulatory profiles (i.e., representing regulatory phonotype) could be generated based on the new and pre-existing mutations on the networks. The profiles will be compared with network operational signatures to predict the outcomes of the newly added mutations. Here mutation has a broad definition: it could be any genetic and epigenetic changes. A network operational signature is a specifically regulatory profile in which a set of genes encodes their regulatory logics and strengths. **Fig 2a,** predictive network models for cancer patients. These models could predict drug targets of tumor clones, prognosis, and drug resistance. The figure shows a case that within a tumor three clones can be identified. Each clone will be modeled independently. The final result of the tumor will be dependent on the outcomes of the three clones. **Fig 2b,** cancer risk predictive network models for healthy individuals. Exome-sequencing data of germline cells will be incorporated into mutational patterns derived from aging, cancer predisposition genes or other cancer-inducing agents. These data will be modeled on hallmark networks to predict whether and when a tumor could occur for a healthy individual.

**Box 1: Mapping cancer hallmark networks to hallmark traits**

**Survival network** represents underlying molecular mechanisms for the most fundamental trait of cancer cells involves their ability to sustain chronic proliferation, resist to cell death, and resist inhibitory signals that might prevent their growth. It represents a collection of three hallmark traits whose underlying biological processes and signaling pathways are highly intertwined and cross-talked.

**Mutating network** represents regulatory mechanisms that trigger high level of genome instability and induce high mutation rate of the genome.

**Dedifferentiation network** represents molecular mechanisms of the capability of unlimited replicative potential of tumors.

**Angiogenesis-inducing network** represents underlying molecular mechanisms for angiogenesis which provides tumors for accessing nutrients and oxygen and evacuating metabolic wastes. It represents stimulating the growth of blood vessels to supply nutrients to tumors.

**Immune-escaping network** represents a mechanism for escaping immune surveillance from host by cancer cells.

**EMT (epithelial-mesenchymal transition) network** represents regulatory machinery that allows cancer cells requiring cell motility ability. Invasion and metastasis has multisteps: cancer cells acquire motility capability, disseminate and circulate in host, and finally colonize in distant organs. EMT network captures the first step of this process. Other steps involved in extensive interactions between cancer cells and tumor microenvironment, which is represented by a stroma-network.

**Genome duplication network** represents a molecular mechanism of triggering genome duplication during cancer clonal evolution. This mechanism has been proposed as a rate-limiting driving force during tumor formation.

**Metabolic network** represents a molecular mechanism of reprogramming energy metabolism which leads to aerobic glycolysis.

**Stroma-network** represents complex interactions between stroma and tumor cells. These interactions could have multiple functions such as tumor-promoting inflammation, immune responses that could eradicate tumors, or other supporting functions for tumor growth.